\documentclass{article}
\usepackage{graphicx}
\usepackage {subcaption} 
\usepackage{color}
\usepackage{enumitem}
\usepackage{tcolorbox}
\tcbuselibrary{skins, breakable, theorems}
\usepackage{multirow}
\usepackage{makecell}
\usepackage[figuresright]{rotating}
\usepackage[title]{appendix}
\usepackage{stackengine}

\usepackage{geometry}
\usepackage{hyperref}
\hypersetup{colorlinks=true,linkcolor=blue,filecolor=red,urlcolor=blue,}

\title{Human game experiment to verify the equilibrium selection controlled by design}
\author{Wang Zhijian, ~ Shan Lixia, ~Yao Qinmei, ~Wang Yijia \\ Experimental social science laboratory, Zhejiang University, Hangzhou, China}

\begin{document}
\maketitle 


\begin{abstract} 
We conducted a laboratory experiment involving human subjects to test the theoretical hypothesis that equilibrium selection can be impacted by manipulating the games dynamics process, by using modern control theory. Our findings indicate that human behavior consists with the predictions derived from evolutionary game theory paradigm. The consistency  is supported by three key observations: (1) the long-term distribution of strategies in the strategy space, (2) the cyclic patterns observed within this space, and (3) the speed of convergence to the selected equilibrium. These findings suggest that the design of controllers aimed at equilibrium selection can indeed achieve their theoretical intended purpose. The location of this study in the knowledge tree of evolutionary game science is presented. 
\end{abstract}

\tableofcontents

\clearpage
\newpage

\section{Introduction}

The primary objective of this experiment study is to validate the consistency between theoretical predictions and experimental results. In our prior theoretical research \cite{zhijian2023nash}, we demonstrated how to manipulate game equilibrium selection using the pole assignment method, which is a sophisticated full-state feedback technique rooted in modern control theory \cite{morse1970,pole2017modern,dorf1995modern}. While previous experimental work has shown how to implement the pole assignment approach for controlling dynamic structures in laboratory experiment involving human subjects \cite{Y5C2022,Behavioral2003}, there has been no exploration of its application to equilibrium selection in such contexts. 

\begin{figure} 
\centering
\begin{tikzpicture}
\tikzstyle{artifact} = [rectangle, draw, align=center, minimum height=1.2cm, minimum width=2.5cm]
\tikzstyle{dep} = [->, thick, >=stealth, line width=2pt, color=black]
\tikzstyle{dep2} = [->,   >=stealth, line width=1.5pt, color=gray]

\node[artifact, fill=cyan!40] (d) at (6,-12) {Differential \\ Equations \\ $\dot{x} =f(x)$};
\node[artifact, fill=cyan!20] (e) at (3.3,-10) {Equilibrium \\ $x^*$};
\node[artifact, fill=cyan!20] (j) at (6.3,-10) {Jacobian \\ $J \big|_{x^*}$};
\node[artifact, fill=green!00] (v) at (9.8,-10.2) {Velocity $\dot{x} \big|_{x}$ 
 \\  \cite{wang2017,wang20115005,2011coyness} };
\node[artifact, fill=cyan!20] (ev) at (3,-8) {Eigenvalue  \\ $\lambda$};
\node[circle, fill=cyan!20] (re1) at (2,-6) {Re};
\node[circle, fill=cyan!20] (im1) at (4,-6) {Im};
\node[circle, fill=cyan!20] (re2) at (8,-6) {Re};
\node[circle, fill=cyan!20] (im2) at (10,-6) {Im};
\node[artifact, fill=cyan!20] (vec) at (8.6,-8) {Eigenvector \\ $\eta$};
\node[artifact, fill=green!00] (stability) at (1,-4) {Stability \\ $\tau$ 
\cite{dan2014,wang2014social,dan2010tasp}};
\node[artifact, fill=green!00] (frequency) at (4,-4) {Angular velocity \\ $\omega$ 
 \cite{wang2014}};
\node[artifact, fill=green!00] (eigencycle) at (11,-4) {Eigencycle \\ $\sigma$
\cite{2021Qinmei,wang2022shujie}};
\node[artifact, fill=green!00] (realcov) at (8,-4) {Coherence  \\ $\kappa$  };

\node[artifact, fill=green!00] (multimode) at (11.5,-8.3) {Multi \\ modes \\ \cite{wang2022mode}};
\node[artifact, fill=green!00] (distrib) at (0,-8.3) {Distribution \\ \cite{2008selten,xu2012test,2015nowak,wang2022shujie,dan2010tasp}};

\node[artifact, fill=green!00] (collapse) at (0.3,-10.6) {Collapse \\ \cite{wang2023pulse}};
\node[artifact, fill=cyan!20] (controller) at (6.3,-2) {Controller};
\node[artifact, fill=green!00] (eq_sel) at (4,0) {Equilibrium   selection\\ \cite{zhijian2023nash} and \textcolor{red}{This study}};
\node[artifact, fill=green!00] (stru_ctrl) at (8,0) {Mode   control \\ \cite{Y5C2022}};

\draw[dep] (d) -- (e);
\draw[dep] (d) -- (j);
\draw[dep] (d) -- (v);
\draw[dep] (j) -- (ev);
\draw[dep] (ev) -- (re1);
\draw[dep] (ev) -- (im1);
\draw[dep] (j) -- (vec);
\draw[dep] (vec) -- (re2);
\draw[dep] (vec) -- (im2);
\draw[dep] (re1) -- (stability);
\draw[dep] (im1) -- (frequency);
\draw[dep] (im2) -- (eigencycle);
\draw[dep] (re2) -- (realcov);

\draw[dep2] (e) -- (distrib);
\draw[dep2] (j) -- (multimode);
\draw[dep2] (d) -- (collapse);
\draw[dep2] (j) -- (controller);
\draw[dep2] (controller) -- (eq_sel);
\draw[dep2] (controller) -- (stru_ctrl);

\end{tikzpicture}
\caption{The location of this study, equilibrium selection (in \textcolor{red}{red text}) by controller, in the knowledge tree of evolutionary game theory and experiment, or called as evolutionary game science. This knowledge tree is presented in the linearization paradigm of dynamics system theory in mathematics. The controller is using pole assignment approach in modern control theory in engineering. The references for related concepts are the existed experimental literatures with associated theory. }
\label{fig:knowledge_tree}
\end{figure}
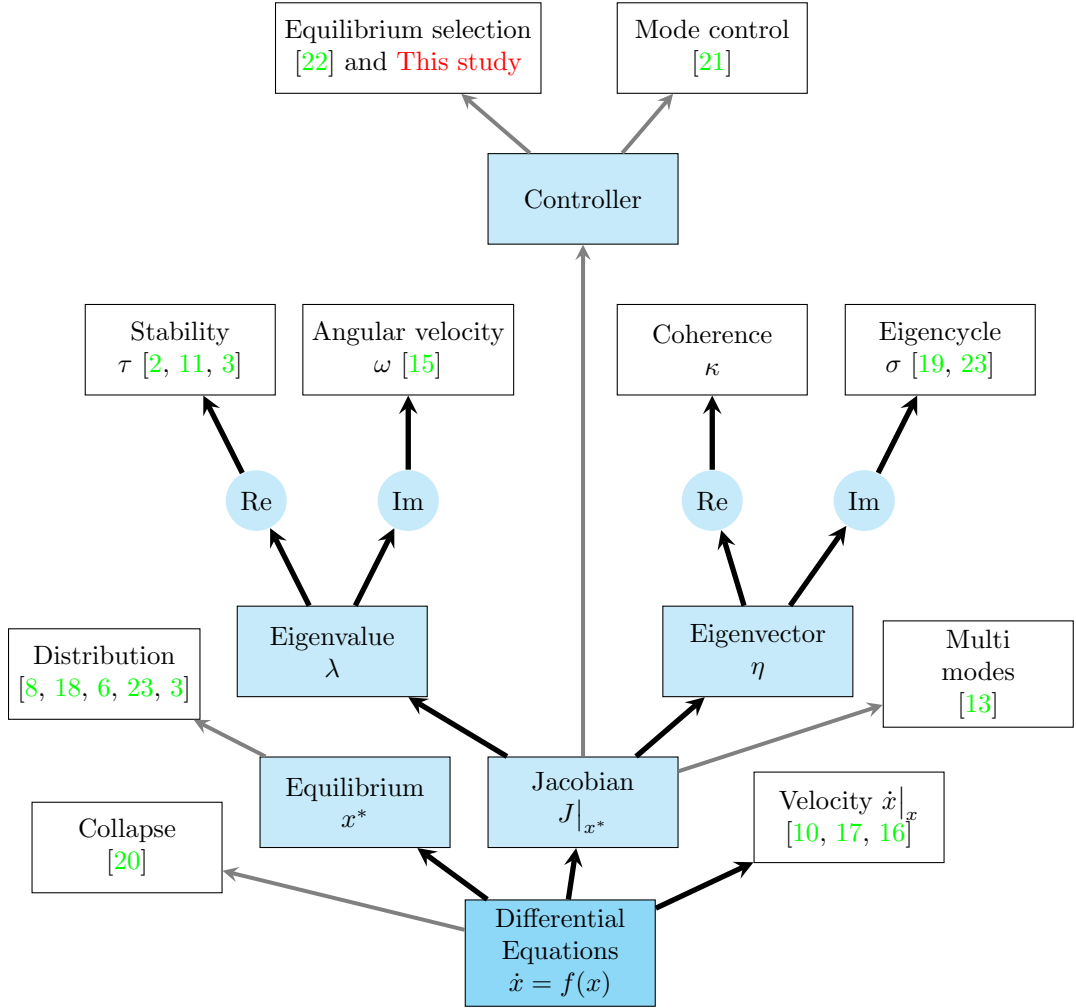

Figure \ref{fig:knowledge_tree} shows the location of this study (in red text), equilibrium selection by controller, in
the knowledge tree of evolutionary game theory and experiment. The tree represents the fundamental concepts of dynamics system theory in mathematics.

This experimental study is based on the previous theoretical work \cite{zhijian2023nash}, where  a five-strategy game was employed to demonstrate the pool assignment method to control the equilibrium selection. In this experimental study, we will utilize the same five-strategy game to realize the control and to verify the consistency.

\subsection{Theoretical background}

\begin{table}[h!]
\caption{The game matrix}
\begin{center}
\begin{tabular}{c|rrrrr}
 \hline 
&x$_1$&x$_2$&x$_3$&x$_4$&x$_5$\\
 \hline 
x$_1$& 0  & 0 &  2 &  0 & -2 \\
x$_2$&2  & 0  & 0  &-2  & 0 \\
x$_3$&0  & 2 &  0 &  2 & -1 \\
x$_4$&$-$2 &  0  & 1 &  0  & 1\\
x$_5$&0  &$-$2 & $-$2 &  1 &  0 \\
 \hline
\end{tabular}
\end{center}
\label{tab:gamemodel}
\end{table}   

\begin{figure}
\centering
\includegraphics[scale=0.45]{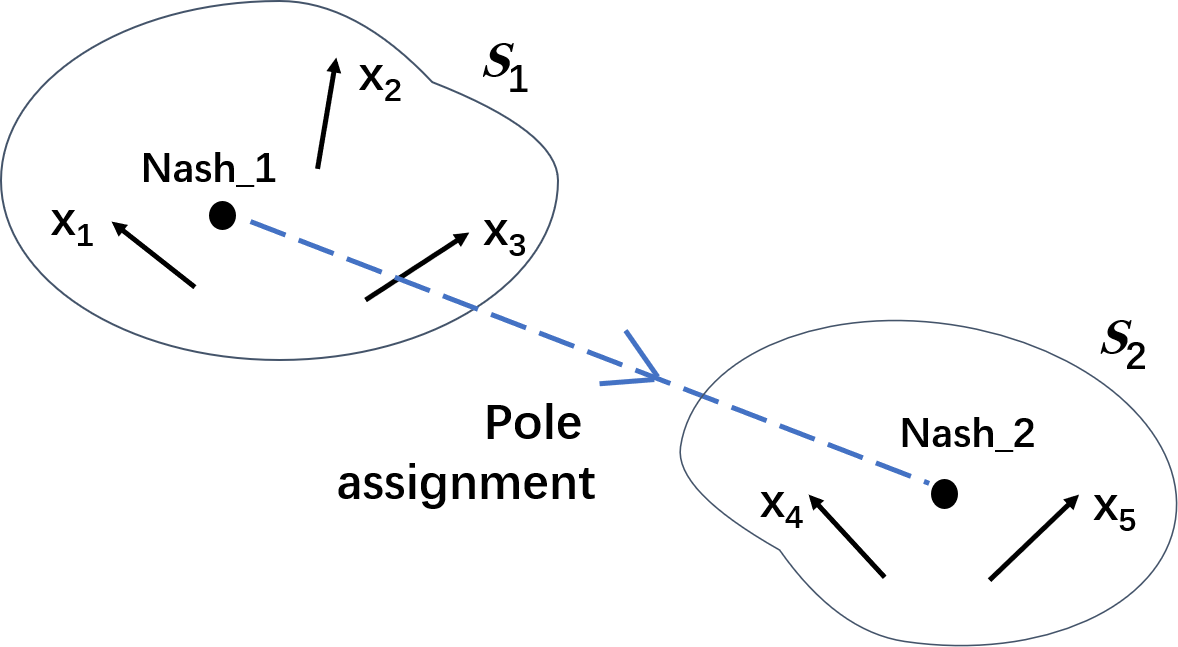} 
\caption{Conceptual figure: 
In a five strategy game, the two equilibrium (Nash\_1 
and Nash\_2) locate in the two sub space $S_1(x_1, x_2, x_3)$ 
and $S_2(x_4, x_5)$, respectively. 
The pole assignment can be designed to make the Nash\_1 
being unstable.
As consequence, the long run trajectory will converge to Nash\_2, 
which means that the equilibrium Nash\_2 is selected.  Alternatively, the pole assignment can be designed to make the Nash\_1 
being more stable, then Nash\_1 will be long rum distribution center. 
This figure comes from \cite{zhijian2023nash}.
\label{fig:concept}}
\end{figure}

The theoretical study \cite{zhijian2023nash} used a symmetric 5-strategy one population game, whose payoff matrix is shown in Table \ref{tab:gamemodel}. The game has and only has two equilibrium (Nash\_1 and Nash\_2) as shown in Fig. \ref{fig:concept}. 
The locations of the two Nash equilibrium of the game are  
\begin{eqnarray}
  {\text{Nash\_1}} = \!\frac{1}{3}(1, 1, 1, 0, 0) \label{eq:Nash1} \\ 
  {\text{Nash\_2}} = \!\frac{1}{2}(0, 0, 0, 1, 1) \label{eq:Nash2}
\end{eqnarray} 
In replicator dynamics \cite{taylor1978evolutionary} as the estimator, the eigenvalues at the Nash\_1 denoted as
$\lambda^o_{\text{Nash\_1}}$ 
are
\begin{equation}\label{eq:lambda_o} 
    \lambda^o_{\text{Nash\_1}} =  \left[
     	\begin{array}{rrrrr}
     	-\frac{1 - \sqrt{3}i}{3} & -\frac{1+\sqrt{3} i}{3}  &  -\frac{2}{3} & -1 & -2 \\
     	\end{array}
     \right].
\end{equation}  
The associated eigenvector, $v_{\text{Nash\_1}} $, which presents the dynamics structure, is as follow.
\begin{equation}\label{eq:eigenvector} 
    v_{\text{Nash\_1}} = \left[ 
     	\begin{array}{rrrrr}
     	 (-.289  & (-.289  & .577 & .151 & -.161 \\
     	   - .5i) &   + .5i) &   &   &   \\
     	 (-.289  & (-.289   & .577 & -.030 & -.462 \\
     	  + .5i) &  - .5i) &   &   &   \\
     	 .577 & .577 & .577 & -.757 & -.221 \\
     	 0 & 0 & 0 & .636 & 0 \\
     	 0 & 0 & 0 & 0 & .844 \\
     	\end{array}
     \right]
\end{equation}      
Notice that, the imaginary part of the first two eigenvalues are not zero, then  the performance of the two associated eigenvectors (1st and 2nd columns) could be a rock-paper-scissors cycle, which is known recently \cite{wang2022shujie,2021Qinmei}  

Utilizing the pole assignment approach, we can manipulate the eigenvalues at the Nash equilibrium point, Nash\_1. In simpler terms, we employ pole assignment to regulate the eigenvalues, which determine the stability of the game evolution, of Nash\_1 by introducing various constant values  $b$ for different experimental treatment. This can be mathematically expressed as follows:
\begin{eqnarray}\label{eq:lambda_c} 
    \lambda^c_{\text{Nash\_1}} &=&\lambda^o_{\text{Nash\_1}} + b~\Big[~1~~1~~ 0~~ 0~~ 0~\Big]   \\
     &=&  \left[
     	\begin{array}{rrrrr}
     	\big(b-\frac{1-\sqrt{3}i}{3}\big) & \big(b - \frac{1+  \sqrt{3}i}{3} \big) &  -\frac{2}{3} & -1 & -2 \label{eq:b}\\ 
     	\end{array}
     \right].
\end{eqnarray}
It is clear that, the original system ($b=0$ condition) is stable at Nash\_1. At the same time, $b=1/3$ is the marginal value for control to switch the stability form stable (when $b<1/3$) to unstable (when $b>1/3$). 

Referring to the control parameter $b$,  the performance of the game system should be difference, which can be derive in dynamics theory and provides a series of theoretical predictions \cite{zhijian2023nash}.

\subsection{Theoretical predictions}
In theory, we have completed the controller design, utilizing the workflow outlined in \cite{zhijian2023nash}. Furthermore, the theoretical predictions of the controller have been validated through numerical agent-based simulations, demonstrating good consistency. In human game experiments, the theoretical predictions remain valid and will be verified.
\begin{enumerate}
    \item When $b < 1/3$, Nash\_1 will be selected equilibrium; when $b > 1/3$, Nash\_2 will be selected equilibrium.
    \item The convergence speed will be faster, when $\Big|b - 1/3\Big|$ being larger.
    \item When When $b < 1/3$, cycles exist constantly; when $b > 1/3$,  cycles do not exist.
\end{enumerate}

\section{Experiment} 
\subsection{Experiment setting}
Our laboratory human game experiments are conducted following the experimental economics protocol, or called as the behaviours game experiment protocol \cite{Behavioral2003}. There are 5 treatments and 40 sessions in our experiments, which are shown in Table \ref{tab:ezpE5c} in the Appendix. 

\paragraph{Game} The game is a symmetry 5-strategy game, which payoff matrix is shown in Table \ref{tab:gamemodel}.  The game was playing in laboratory using web base application.  It is a  repeated game, each player plays the game with all of the other participate at every round.  

\paragraph{Treatment}  For various treatments, the game is invariant, but the treatment parameter, $b$, is different. 
That is, the treatments identified by $b$, which control the eigenvalues of a Nash equilibrium (Nash\_1 in Eq. \ref{eq:Nash1}).  
There are 5 treatments including 40 sessions experiments in total. The mapping between the treatment ID and the control parameter $b$ is shown in Table \ref{tab:ezpE5c} in the Appendix. 

\paragraph{Subjects}  There are 30 university students participating the games. The 30 students are divided to 6 groups, and each includes 5 subjects. 
Each group of 5 subjects, in 2 hours, will play the game 8 sessions. In these 8 sessions, the 5 treatments are assigned by experimenter. 

\paragraph{Process} The 5 student subjects play the games and make decision which of the 5 strategy to choose at each descrete rounds. Each round last about 2 second. Each session play 360 rounds, or 720 seconds, or 12 minutes.  Between two successive sessions, there were 3 minutes rest time for subjects, at the same time, for the experimenter to set the parameters for the coming session.   

\paragraph{Payment} The average pay of the subjects for the 2 hours game experiment is 150 Yuan RMB. They pay as 50, 100, 150, 200 or 250 Yuan, respectively, referring to the rank of the total score gain in the 8 sessions in the games. The higher of the score, the pay is more. 

\paragraph{Experiment time} The experiments are conducted during February and Match in 2023. The organization of the sessions, treatments and related settings 
are list in Table \ref{tab:ezpE5c} in the Appendix. 

\subsection{Controller realization } 
The pool assignment controller has been realized in agent-based simulation and human game experiment \cite{zhijian2023nash,Y5C2022}
to control the evolutionary dynamics process to influence the game outcome. The controller is realize in two level, treatment level (environment) and round level (real time level). 

\paragraph{Treatment level} In theory, the control matrix $K$ calculate as follow steps.
\begin{itemize}
    \item set the channel B = [~0~0~0~1~1~] as the constant for all treatments. 
    \item set the treatment parameter $b$, then the disired eigenvalues, $\lambda^c_{\text{Nash\_1}}$, can be obtain by Eq. (\ref{eq:b}) 
    \item calculate the feedback matrix $K$, with $\lambda^c_{\text{Nash\_1}}$, $B$ and the Jacobian at Nash\_1.  
\end{itemize}
As the results, we set $b=-0.8, -0.4, 0, 0.4, 0.8$ as the treatment parameter, then $K$ values, which is a $1\times 5$ matrix (or a vector), are list in rows in the follow Table \ref{tab:Kvalue}

\begin{table}[!ht]
    \centering
    \begin{tabular}{c|r|r|r|r|r}
    \hline 
~~~~$b$~~~~&$k_1$~~~~&$k_2$~~~~&$k_3$~~~~&$k_4$~~~~&$k_5$~~~~\\\hline
 $-$0.8&0.5247 & ~0.9485 & $-$1.4732 & $-$1.8335 & ~0.2335 \\
 $-$0.4&0.3834 & 0.2623 & $-$0.6458 & $-$0.8476 & 0.0476 \\
0& 0 & 0 & 0 & 0 & 0 \\
0.4& $-$0.6256 & 0.1614 & 0.4641 & 0.7092 & 0.0908 \\
 0.8&$-$1.4933 & 0.7467 & 0.7467 & 1.2800 & 0.3200 \\\hline
    \end{tabular}
    \caption{Controller gain matrix $K$'s value when ${B} = \big[0~ 0~ 0~ 1~ 1\big]^T$, and data from  \cite{zhijian2023nash}}
    \label{tab:Kvalue}
\end{table}

\paragraph{Round level} 
In experiment, the controller works after all of the player making the decision. The controller is realized by computer calculate. The steps are as follow:
\begin{enumerate}
    \item Estimate the social state $x$, the distribution of the 5 subjects strategy. For example, $x=[3,2,0,0,0]$ means 3 subjects choose strategy x$_1$, 2 subjects choose strategy x$_2$, and none choose strategy x$_3$, x$_4$, or x$_5$.
    \item Calculate the payoff of the game, reward and tax for each subject, referring to the strategies that the subjects choose. The algorithm for these three term are given in \cite{Y5C2022,zhijian2023nash}.  
    \item Provide the feedback information for each subject, including its own strategy, social state (strategy distribution of the 5 subjects). its own game payoff, reward, tax and their sum as the total earn points at this round.   
\end{enumerate}
The algorithm in 2-point is the central tech to realize the controller at round level (real time level). 

\section{Verification}

In this report, the three key observations are employed to verify the consistency. They are (1) the long-term distribution of strategies in a high-dimensional strategy space, (2) the cyclic patterns observed within this space, and (3) the speed of convergence to the selected equilibrium under controlled conditions.

We introduce the measurements and the results of theory and experiment comparisons in this section. The theoretical result comes from \cite{zhijian2023nash}. 

\subsection{Measurements}\label{subsec:measurement} 
Comparing to the original system, the controlled system has observable consequence.  
The verifiable theoretical predictions of the controller, as well as their measurements, are listed as follow.  

\begin{enumerate}
\item  \textbf{Distribution.}  by changing the 
pole assignmant parameter $b$. The strategy distribution of 
the controlled game should be identical to that of the original game. 
The theoretical expectation is 
\begin{eqnarray}
\rho\!&\rightarrow&  {\text{Nash\_1}} = \!\frac{1}{3}(1, 1, 1, 0, 0)~\text{~~~~when~}~ b  \rightarrow -1, \label{eq:toNash1} \\ 
   \rho\!&\rightarrow&  {\text{Nash\_2}} = \!\frac{1}{2}(0, 0, 0, 1, 1)~\text{~~~~when~}~ b \!\rightarrow\!1.  \label{eq:toNash2}
\end{eqnarray} 


In measurement, the prediction of equilibrium selection, the theoretical distribution $\rho^T$ from computer agent based simulation,  can be 
verified by the experimental distribution $\rho^E$ from human experiment. $\rho$ is the strategy proposition vector form the time series. In formulae, 
  \begin{eqnarray}\label{eq:meanrho}
\bar{\rho}_{i} = \frac{1}{T}\sum_{t=0}^T \rho_{i}(t),  
\end{eqnarray}  
where $\rho_i(t)$ is 
the proportion of $i$-th strategy used at time $t \in [0, T]$. For long time, the beginning phase effect can be ignore. In our experiment report, $\bar{\rho^E}(b)$ is the average over $T = [1,2,...~360]$ in a session, then average over all of the sessions in a given treatment $b$. The results are shown in Figure \ref{fig:distri_result}.

\item \textbf{Converge speed.} The converge speed to desired equilibrium is impacted by the pole assignment parameter, $b$. 
The convergence speed is defined as the time cost of 
time from  the full randomly 
initial distribution [1, 1, 1, 1, 1]/5 to equilibrium 
(Nash\_1 or Nash\_2). 

In measurement,  we can measure the time dependent 
Euclidean distance $d(t)$ from Nash\_1 and 
 Nash\_2 respectively, 
\begin{eqnarray}
    d_{\text{Nash\_1}}(t) &=& \big|\rho(t) -  {\text{Nash\_1}}\big| ~\text{~~~~when~}~ b  < 1/3, \\
    d_{\text{\text{Nash\_2}}}(t) &=& \big|\rho(t) -  {\text{Nash\_2}}\big| ~\text{~~~~when~}~ b  >  1/3. 
\end{eqnarray} 
Here, shown in Eq \ref{eq:b}, the marginal value for the stability of Nash\_1 is -1/3, then $b>1/3$ is the marginal value for control. 
Then, if the Nash\_1 is selected, the system converge to Nash\_1, $d_{\text{Nash\_1}}(t)$ will close to 0. If convergence speed is faster, $d_{\text{Nash\_1}}(t)$ declines to 0 is faster. Alternatively,   if the Nash\_2 is selected, the system converge to Nash\_2, $d_{\text{Nash\_2}}(t)$ will close to 0. If convergence speed is faster, $d_{\text{Nash\_2}}(t)$ declines to 0 is faster. 

\item \textbf{Cycle.} In theory, the eigencycles 
in the 2-d subspace of the game can be calculated from 
the complex eigenvector associated to the eigenvalue shown in Eq. \ref{eq:lambda_o}, 
 referring to \cite{wang2022shujie}.   
This can be verified by  time average of the angular momentum 
measurement along the time series, referring to \cite{wang2022shujie}, 
\begin{eqnarray}\label{eq:meanL}
\bar{L}_{mn}&= & \frac{1}{t'}\sum_{t=0}^{t'} x_{mn}(t) \times x_{mn}(t + 1) 
\end{eqnarray}  
Herein, $x_{mn}(t)$ is the strategy vector in the $mn$ subspace 
(2-d subspace, $m>n$ and $m,n \in [1,2,3,4,5]$), $t'$ is the length of time series. 
For a 5 strategy game, the identical 2-d subspace number is 10, 
and the observer sample is 10 \cite{wang2022shujie,2021Qinmei,WY2020}. 
The equivalent of the theoretical eigencycle and the observed angular momentum $L$ in is proved referring to 
\cite{2021Qinmei,WY2020}.

In measurement, the cycle strength $|\bar{L}|$ is defined as 
\begin{equation}\label{eq:absL}
    |\bar{L}| = \big(\sum_{mn} L_{mn}^2\big)^{1/2} 
\end{equation}  
Herein, the subscript $mn$ are the index of the dimension of the 5-strategy strategy space \cite{WY2020,wang2022shujie,zhijian2023nash}.  the in the time series to verify the theoretical expectation of the strength of the cycles.    

\end{enumerate}
 
\subsection{Results}
This section reports the experiment and the verifications results, which are illustrated by the Figure \ref{fig:distri_result},  Figure \ref{fig:d_t} and  Figure \ref{fig:cycle_absL}.
\subsubsection{Distribution}

\begin{figure}[ht!]
\centering
\includegraphics[scale=.35]{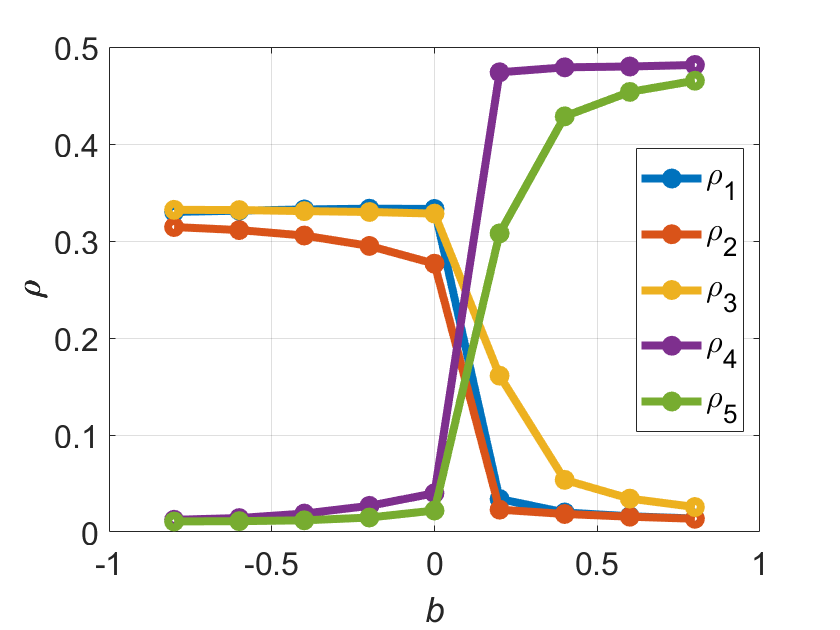} \\
\includegraphics[scale=.35]{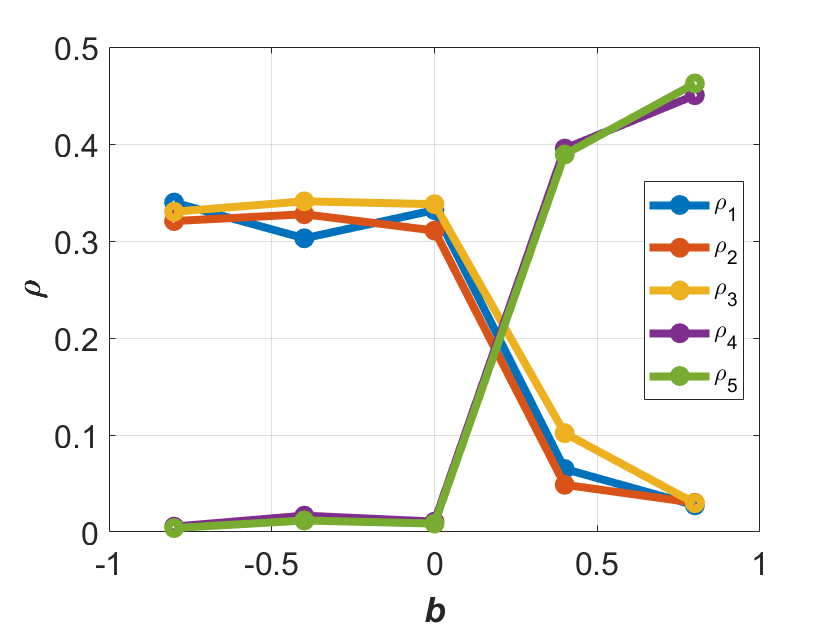} 
\caption{\label{fig:distri_result} 
 (\textbf{a}) Theory and (\textbf{b}) Experiment. Distribution of the long run average 
$\rho(b)$. 
when $b  \rightarrow -0.8$, 
the trend is Nash\_1 selected, and the stategy x$_4$, x$_5$ is dominated; 
alternatively, when $b \rightarrow 0.8$, 
 Nash\_2 selected, and the stategy x$_1$, x$_2$, x$_3$ is dominated. The consistency of the theory and experiment performs well.} 
\end{figure}

The consistency between the theory and the experiment is obviously. Figure \ref{fig:distri_result} shows the distribution of the long run average 
($\bar{\rho_1},\bar{\rho_2},
\bar{\rho_3},\bar{\rho_4},\bar{\rho_5}$)
as the function of the pole assignment parameter $b$.

\paragraph{Theory} 
Figure \ref{fig:distri_result}(a)  These results comes from abed simulations. When $b$ being $ 0 \rightarrow -1$, 
the trend is to select 
 equilibrium Nash\_1, where $\rho_1, \rho_2, \rho_3$ are larger which means the strategy x$_1$, x$_2$, x$_3$ are domination; 
Alternatively, when $b$ being $ 0 \rightarrow 1$, 
the trend is to select equilibrium Nash\_2, where $\rho_4, \rho_5$ are larger which means the strategy x$_4$, x$_5$ are domination.

\paragraph{Experiment} 
Figure \ref{fig:distri_result}(b) shows the human subject game experimental results. $\rho^E (b)$ present the time average of 
the proportion of $i$-th strategy used in the experiment sessions for each treatment identified by the control parameter $b$.  

\paragraph{Consistency} Obviously, as shown in Figure \ref{fig:distri_result}, the results from human subject game experiment consist  with the theoretical results well.

\subsubsection{Converge speed}

\begin{figure}[ht!]
\centering
\includegraphics[scale=.57]{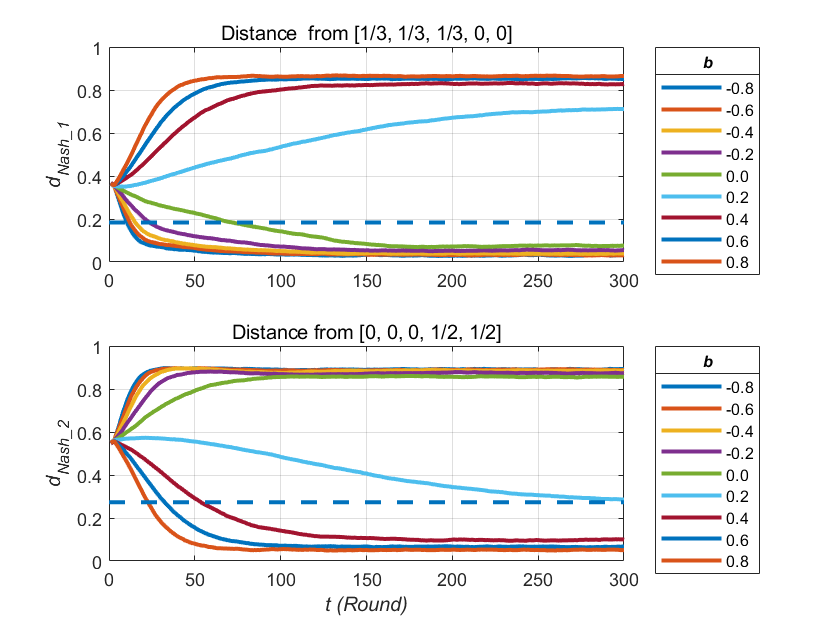}  
\includegraphics[scale=.57]{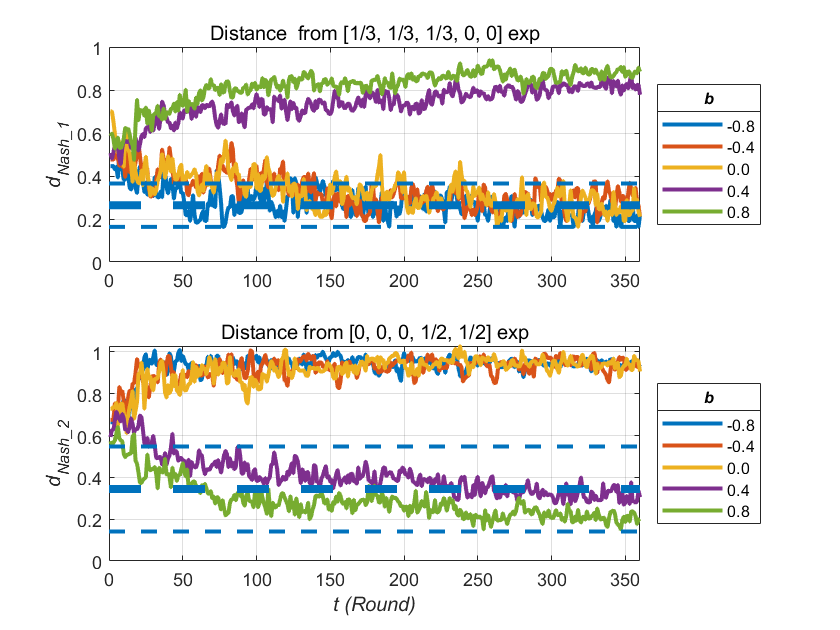}  
\caption{\label{fig:d_t} 
 (\textbf{a}) Theoretical $d^T_{\text{Nash\_1}}(t)$.  (\textbf{b}) Theoretical $d^T_{\text{Nash\_2}}(t)$.  (\textbf{c}) Experimental $d^E_{\text{Nash\_1}}(t)$.  (\textbf{d}) Experimental $d^E_{\text{Nash\_2}}(t)$. Obviously, convergence speed to Nash\_1 ([1/3, 1/3, 1/3, 0, 0]) is the fastest when $b=-0.8$. Alternatively, convergence speed to Nash\_2 ([0, 0, 0, 1/2, 1/2]) is the fastest when $b=+0.8$. The consistency of the theory and experiment performs well.  }  
\end{figure}

On the converge speeds to related Nash equilibrium, the consistency between the theory and the experiment is obviously, too. Figure \ref{fig:d_t} shows $d_{\text{Nash\_1}}(t)$  and $d_{\text{Nash\_1}}(t)$ for each of the parameters $b$. 

\paragraph{Theory}  
Showing in Fgiure \ref{fig:d_t} (a), in $b=-0.8, -0.4, 0$ treatments, the social state closing Nash\_1, as $d_{\text{Nash\_1}}$ decline to zero along time. In these three treatments, $b=-0.8$ closing to its zero (the Nash equilibrium, Nash\_1, [1/3, 1/3, 1/3, 0, 0]) is the fastest.  

Showing in Fgiure \ref{fig:d_t} (b),  Alternatively, in $b= 0.4, 0.8$ treatments, the social state closing Nash\_2, as $d_{\text{Nash\_2}}$ decline to zero along time. In these two treatments, the $b= 0.8$ closing to its zero (the Nash equilibrium, Nash\_2, [0, 0, 1/2, 1/2, 1/2]) is faster. 

\paragraph{Experiment} 
As shown in Fgiure \ref{fig:d_t} (c), in $b=-0.8, -0.4, 0$ treatments, the social state closing Nash\_1, as $d^E_{\text{Nash\_1}}$ decline to zero along time. In these three treatments, $b=-0.8$ closing to its zero (the Nash equilibrium, Nash\_1, [1/3, 1/3, 1/3, 0, 0]) is the fastest.  

As shown in Fgiure \ref{fig:d_t} (d),  alternatively, in $b= 0.4, 0.8$ treatments, the social state closing Nash\_2, as $d^E_{\text{Nash\_2}}$ decline to zero along time. In these two treatments, the $b= 0.8$ closing to its zero (the Nash equilibrium, Nash\_2, [0, 0, 1/2, 1/2, 1/2]) is faster. 

\paragraph{Consistency} Obviously, as Shown in Figure \ref{fig:d_t}, the results from human subject game experiment consist with the theoretical results well, too.

\subsubsection{Cycles}

\begin{figure}[ht!]
\centering 
 \includegraphics[scale=.35]{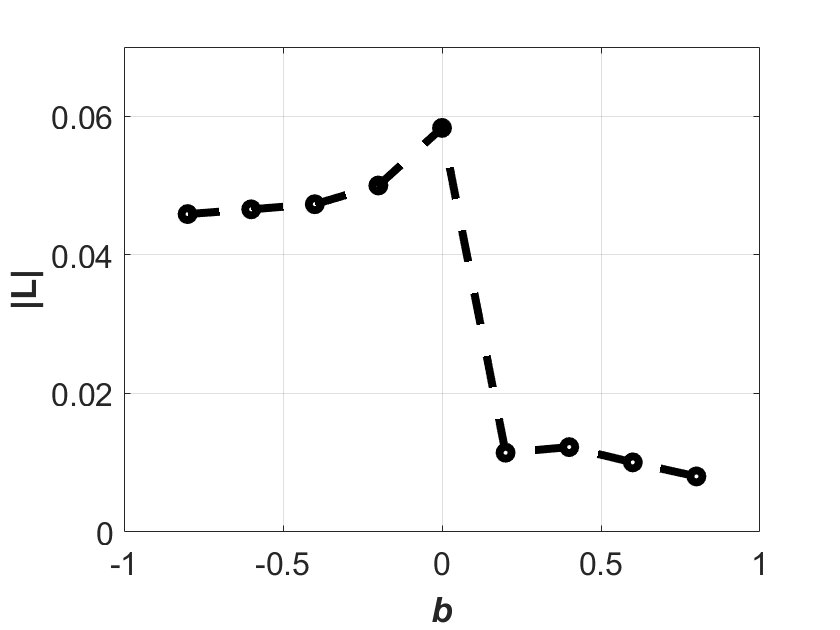} \\
 \includegraphics[scale=.35]{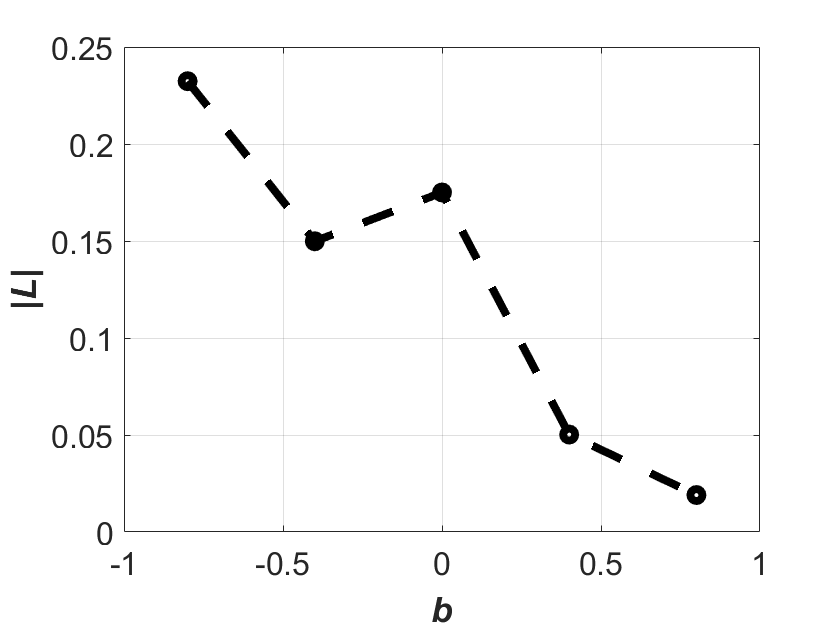} 
\caption{\label{fig:cycle_absL}  
 (Top, \textbf{a}) theoretical cycle strength $|L^T|$ as a function of $b$.  (Bottom, \textbf{b}) experimental cycle strength $|L^E|$  as a function of the parameter $b$. Obviously, the cycles exist constantly when $b = -0.8, -0.4, 0$; alternatively, the cycle disappears when $b = +0.4, +0.8$. The consistency of the theory and experiment performs well. } 
\end{figure}

On cycles, the consistency between the theory and the experiment is obviously, too. Figure  \ref{fig:cycle_absL} shows  cycle strength $|L(b)|$, referring to Eq \ref{eq:absL}, as a function of the parameters $b$.

\paragraph{Theory} Figure \ref{fig:cycle_absL}(a) shows the  cycle strength $|L|$. 
In theory, when the equilibrium remained 
at Nash\_1 the cycle in the $(x_1, x_2, x_3)$ subspace is significant; 
Alternatively, when the equilibrium shift to the Nash\_2 equilibrium, the cycles disappear, and $|L| \rightarrow 0$ when $b = 0.4, 0.8$. 

 \paragraph{Experiment} 
Figure \ref{fig:cycle_absL}(b) shows the experimental results on the cycle strength. The pattern is the same as the theoretical predictions. 

\paragraph{Consistency} Obviously, as shown in Figure \ref{fig:cycle_absL}, the results from human subject game experiment consist with the theoretical results well, again.

\section{Discussion}
\subsection{Summary}

  The consistency of the theory and experiment, in our data, is supported obviously by the three key observations: (1) the long-term distribution of strategies in a high-dimensional strategy space, (2) the cyclic patterns observed within this space, and (3) the speed of convergence to the selected equilibrium under controlled conditions. These observations suggest that the design of controllers aimed at selecting Nash equilibria can indeed achieve their theoretical intended purpose. 

We hope, these enhance our understanding of evolutionary game dynamics theory and experiments, particularly in the area of dynamics process control in strategy interactions in games.

\subsection{Related works}
\paragraph{Presenting this study in the knowledge tree (Figure \ref{fig:knowledge_tree}):} Last decade has seen several significant development on the consistency of the game dynamics theory and experiment, like identifying cycle pattern on rock-paper-scissors game \cite{dan2014,wang2014social}, 
cycle frequency on 2 x 2 game \cite{wang2014}, dynamics structure in high dimensional games \cite{wang2022shujie, 2021Qinmei,wang2022mode}. Recently, extends to control the dynamics structure \cite{Y5C2022}. In this paper, to our knowledge, the experimental verification of the control of the game equilibrium selection is the first.  Figure \ref{fig:knowledge_tree} illustrates the location of this work in the knowledge tree for the evolutionary game theory and experiment.

\paragraph{Regarding Control of Equilibrium Selection:} The theoretical expectation, as outlined in \cite{zhijian2023nash}, is that adjusting the control parameter $b$ will alter the eigenvalue at the Nash equilibrium, thereby modifying its stability. Consequently, this change can influence the long-run distribution. Additionally, varying $b$ affects the dynamics structure and convergence speed. The consistency between theoretical predictions and experimental results is excellent in this context.
\paragraph{Regarding Control of Dynamics Structure (Mode):} Previous experiments have demonstrated, as reported in \cite{Y5C2022}, the effectiveness of the pool assignment full state feedback approach. This method allows for the control of the dynamics structure (mode) in high-dimensional space to achieve desired outcomes. The consistency between theoretical predictions and experimental results is also excellent in this context.

\paragraph{Regarding equilibrium selection:}
Equilibrium selection is an important issue in game theory \cite{1988harsanyiSelten}. It relates to the long run outcome of the strategy interaction of a social system. To uor knowledge, this experimental study together with it theoretical analysis \cite{zhijian2023nash} provides a new approach to influence the equilibrium selection to desired goal. We hope it is a helpful adding knowledge to game theory.



\clearpage
\section{Appendix: Human game experiment protocol}\label{sec:human}

\subsection{Instructions for subjects (example)}
You are participating in a social science experiment. You will earn corresponding RMB based on your and your opponents' decisions. Therefore, it is important to read these instructions carefully. The instructions we provide are for your personal use only. If you have any questions, please raise your hand.

~\\
In the experiment, each participant's rules are exactly the same.  

~\\
There are a total of 8 sessions, each lasting 12 minutes. In the experiment, you will play multiple rounds with all opponents.  

~\\
The screen shots of the game playing are shown in  Figure \ref{fig:sc1}, Figure \ref{fig:sc2} and Figure \ref{fig:sc3}. 

~\\
In each round, you and your opponent each choose a strategy (see point A in the figure) and do not know each other's strategy in advance. After each round, you will see your own and the other four opponents' strategies (in [~~~~~] brackets, in order representing the number of players for each strategy), as well as your own payoff (see point B in the figure). The payoff includes our own game score (according to the payment matrix), taxes (Tax), and subsidies (Reward). The sum of our game score, taxes, and subsidies equals the final score for that round. The rules for subsidies and taxes depend on the choices of all five people, and the rules for each session are fixed.\\

~\\
In this experiment, your ultimate goal is to maximize your score. At the end of the experiment, we will determine the experiment reward based on your total score rank among group members. Taking 5 people as an example, your cumulative score (see point C in the figure) (the cumulative score is obtained by adding up the "final score for that round" for each round) will be ranked with your other 4 opponents, and the reward will be paid according to the ranking from high to low. The first place will receive 250 RMB, the second place 200 RMB, decreasing by 50 RMB each time, and the last place will receive 50 RMB. In the case of tied scores, the average of the two scores will be paid. The reward will be paid promptly after the experiment is over.\\

~\\
Experimental steps: At the beginning of the experiment, please open the experiment website 10.10.70.xx, click on the experiment program, enter the experiment code and your personal code, and enter the experimental interface. Click the button to select your strategy for this round, and submit it by pressing the confirm button. After each submission, the left side of the page will display the strategies of both parties in the last round, your score, the current round number, and your cumulative score.\\

~\\
After the experiment is over, we will collect this experimental manual. Thank you for your cooperation!\\

\begin{figure}[!ht]
\centering
 \includegraphics[scale=0.35]{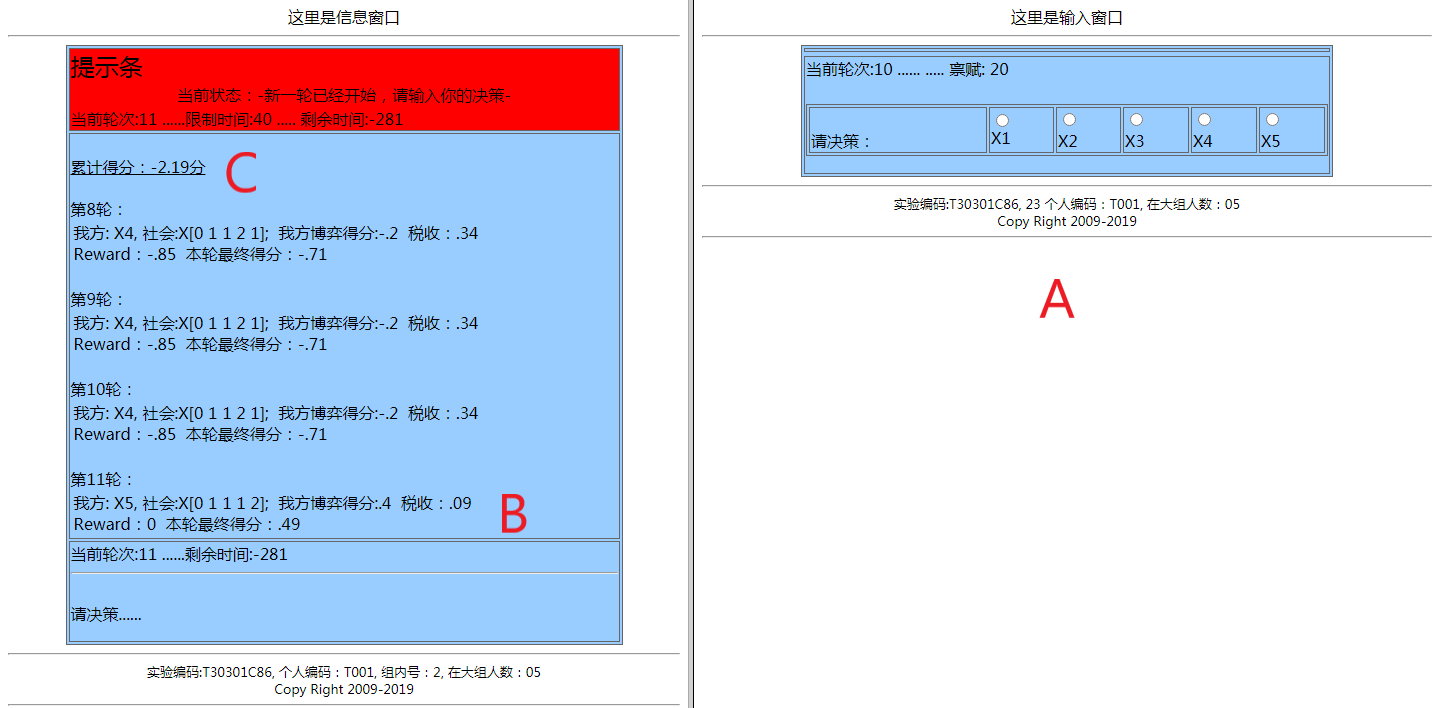} 
 \label{fig:sc4} 
\caption{The screen shot for decision making. A zone is the decision making, where a subject can choose one of the 5 strategy of the game. B zone is the results of the last round, where the own strategy, social state, game earn, reward and tax, as well as the sum score of the round are reported. C zone is the accumulated total score of the session of the subject. } 
\end{figure}

\subsection{Experiment processes}

The process is the same as that of the repeated games experiment involving human subject in laboratory, like \cite{dan2014,wang2014social,Behavioral2003}, in experimental economics. The incentive of the student subject is the payment in current referring to the total score they won in the repeated games.

\paragraph{Student subjects}
Some character  of our experiment can be seen as the screen shot as follow.
\begin{enumerate}
    \item  
As the experiment begins, the following information and input window will be on the computer screen:\\
\begin{figure}[!ht]
\centering
 \includegraphics[scale=0.35]{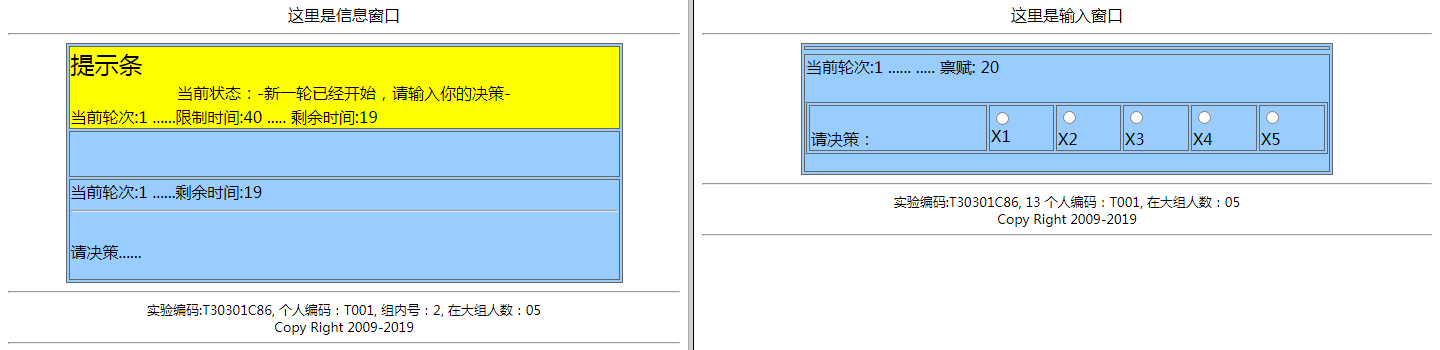} 
  
\caption{\label{fig:sc1} As the experiment begins, the following information and input window will be on your computer screen.} 
\end{figure}

    \item 
After one round, you will observe the following information window and input window:\\
\begin{figure}[!ht]
\centering
 \includegraphics[scale=0.35]{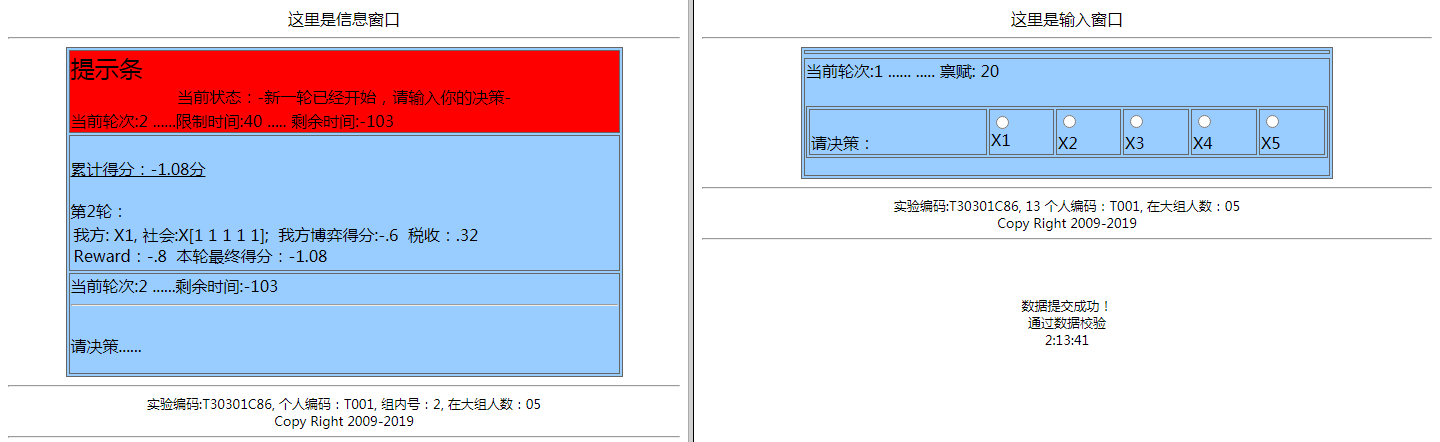} 
\caption{\label{fig:sc2}After one round, you will observe the following information window and input window.} 
\end{figure}

    \item 
Using F5 to refresh, the right input window will then list your options X1 and X2 and X3 and X4 and X5 (as below):\\
\item 
The game is repeated about 360 rounds in a experiment session. There are total 8 sessions in two hours for the subjects to participate.
\item 
The subjects got their payment referring to their own score, respectively, and the instructions for subjects.
\begin{figure}[!ht]
\centering
\includegraphics[scale=0.35]{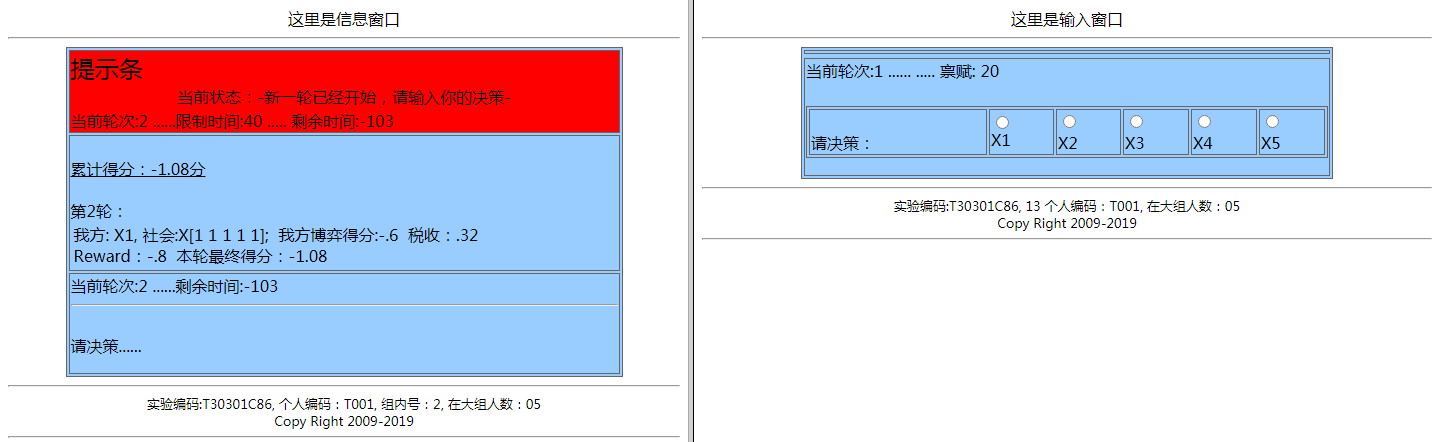} 
\caption{\label{fig:sc3} Using F5 to refresh, the right input window will then list your options X1 and X2 and X3 and X4 and X5.} 
\end{figure}
\end{enumerate}

\subsection{Experiment organisation}

There are 5 treatments and 40 sessions in our experiments, which are shown in Table \ref{tab:ezpE5c} in the Appendix. 

\paragraph{Treatments} The first column indicates the treatments,. Where the two digits indicate the control parameter $b$ values. In details, 
\begin{itemize}
    \item N2 means $b=-0.8$ 
    \item N1 means $b=-0.4$
    \item ~o~ means $b=0$ 
    \item P1 means $b=~0.4$ 
    \item P2 means $b=~0.8$ 
\end{itemize}

\begin{table} 
\caption{The experiment session organisation}
\begin{center}
\begin{tabular}{|c|ccccc|c|}
 \hline 
Treatment &SessionID&	Date& Subjects&Period& Permutation & $b$ \\
 \hline  
N2& 30217A86N21&	20230217&	5&	360&	14 & $-$0.8\\
  & 30217B86N21&	20230217&	5&	360&	14 & \\
  & 30217C86N21&	20230217&	5&	360&	14 & \\
  & 30227A86N21&	20230227&	5&	360&	25 & \\
  & 30227A86N22&	20230227&	5&	360&	35 & \\
  & 30227B86N21&	20230227&	5&	360&	25 & \\
  & 30314A86N21&	20230314&	5&	360&	34 & \\
  & 30314A86N22&	20230314&	5&	360&	14 & \\
\hline
N1& 30217A86N11&	20230217&	5&	360&	25 & $-$0.4\\
  & 30217B86N11&	20230217&	5&	360&	25 & \\
  & 30217C86N11&	20230217&	5&	360&	25 & \\
  & 30227A86N11&	20230227&	5&	360&	15 & \\
  & 30227B86N11&	20230227&	5&	360&	15 & \\
  & 30227B86N12&	20230227&	5&	360&	35 & \\  & 30314A86N11&	20230314&	5&	360&	25 & \\
&30314A86N12&	20230314&	5&	360&	00 & \\
\hline
o & 30217A86o11&	20230217&	5&	360&	15 & 0\\
  & 30217A86o12&	20230217&	5&	360&	35 & \\
  & 30217B86o11&	20230217&	5&	360&	15 & \\
  & 30217B86o12&	20230217&	5&	360&	35 & \\
  & 30217C86o11&	20230217&	5&	360&	15 & \\
  & 30217C86o12&	20230217&	5&	360&	35 & \\
  & 30314A86o11&	20230314&	5&	360&	15 & \\
  & 30314A86o12&	20230314&	5&	360&	35 & \\
\hline
P1& 30217A88P11&	20230217&	5&	360&	24 & 0.4\\
  & 30217A88P12&	20230217&	5&	360&	00 & \\
  & 30217B88P11&	20230217&	5&	360&	24 & \\
  & 30217B88P12&	20230217&	5&	360&	00 & \\
  & 30217C88P11&	20230217&	5&	360&	24 & \\
  & 30217C88P12&	20230217&	5&	360&	00 & \\
  & 30227A88P11&	20230227&	5&	360&	24 & \\
  & 30227A88P12&	20230227&	5&	360&	00 & \\
  & 30227B88P11&	20230227&	5&	360&	24 & \\
  & 30227B88P12&	20230227&	5&	360&	00 & \\
  & 30227B88P13&	20230227&	5&	360&	14 & \\
  & 30314A88P11&	20230314&	5&	360&	24 & \\  
\hline
P2& 30217A88P21&	20230217&	5&	360&	14 & 0.8\\
  & 30217A88P22&	20230217&	5&	360&	34 & \\
  & 30217B88P21&	20230217&	5&	360&	14 & \\
  & 30217B88P22&	20230217&	5&	360&	34 & \\
  & 30217C88P21&	20230217&	5&	360&	14 & \\
  & 30217C88P22&	20230217&	5&	360&	34 & \\
  & 30227A88P21&	20230227&	5&	360&	14 & \\
  & 30227A88P22&	20230227&	5&	360&	34 & \\
  & 30227A88P23&	20230227&	5&	360&	14 & \\
  & 30227B88P21&	20230227&	5&	360&	14 & \\
  & 30227B88P22&	20230227&	5&	360&	34 & \\ 
  & 30314A88P21&	20230314&	5&	360&	14 & \\   
 \hline
\end{tabular}
\end{center}
\label{tab:ezpE5c}
\end{table}

\paragraph{ExperimentID} 
The second column, ExperimentID, represents a single experimental session. The 13 digits, for example "30217A88P2114", can be divided into 8 group of data: experiment date ("30217"), time period ("A"), server code ("88"), control parameter symbol ("P"), control parameter intensity ("2"), repetition number within the time period ("1"). 
Following is the description of each column:
\begin{enumerate}
    \item 
Experiment date: represented as a 5-digit integer in the format YMMDD.
    \item 
Time: a single character string indicating whether the experiment was conducted in the morning (A), afternoon (B), or evening (C).
    \item 
Server code: a two-digit integer representing the last two digits of the server IP address.
    \item 
Control parameter symbol: a single character indicating the sign of the control parameter. P represents a positive control parameter, o represents a control parameter value of zero, and N represents a negative control parameter.
    \item 
Control parameter intensity: an integer indicating the strength of the control parameter, corresponding to a multiple of 0.4 in the main text. In this experiment, the values are 0, 1, or 2.
    \item 
Repetition number: an integer indicating the order in which experiments with the same control parameter were repeated.

\end{enumerate}

\paragraph{Date} The third column is the experiment date in YYYMMDD format, or year-month-day.

\paragraph{Subjects} The fourth column is the number of student human subject participated the experiment. 

\paragraph{Period} The fifth column is the number of repeated periods of the game in the experiment. In data analysis, we use of the first 360 rounds. 

\paragraph{Permutation} The sixth column is the the index ($ij$) of the permutation between the strategy $i$ and strategy $j$. 
 
To avoid the influence of prior experience on participants between consecutive experimental sessions, we adopted a method of permutation two out of the five strategies in the game matrix before each session.

This was due to the fact that the experiment had exactly only two Nash equilibria, 
\begin{itemize}
    \item Nash\_1 (1/3,1/3,1/3,0,0) and 
    \item Nash\_2 (0,0,0,1/2,1/2).
\end{itemize}
 The experimental results showed that participants either chose strategies 1, 2, and 3 or strategies 4 and 5. 

Each experiment consisted of 8 sessions, with 360 rounds per session. If the permutation method was not employed among these 8 sessions, many sessions would converge to the same equilibrium point, allowing participants' memories from previous sessions to influence their choice of game strategies in the current session.

Using the initial payoff matrix resulted in relatively monotonous experimental outcomes, and participants tended to make choices influenced by the results of the previous session. To break the monotony of the experimental results and enhance flexibility, we made the following strategic adjustments to the initial payoff matrix to one of the permutation as follow: 
\begin{itemize}
    \item swapping strategy 1 with strategy 4, 
    \item swapping strategy 1 with strategy 5, 
    \item swapping strategy 2 with strategy 4, 
    \item swapping strategy 2 with strategy 5,
    \item swapping strategy 3 with strategy 4, 
    \item swapping strategy 3 with strategy 5.
\end{itemize}

Due to the permutation of strategies in the game matrix, the corresponding control channel vector $B$ and gain matrix vector $K$ also needed to be permuted accordingly. Furthermore, adjustments were made when calculating game payoffs, reward, and taxes to ensure that the system remained consistent after the strategy permutation.

\paragraph{$b$} The last column is the  parameter $b$, which modifying the eigenvalue, that relates to the stability of the game system at the Nash\_1. Its related control parameter $K$ is calculated by pool assignment approach.

\bibliographystyle{plain}

\end{document}